\documentclass[aps,preprint,showpacs,preprintnumbers,amsmath,amssymb,floatfix]{revtex4}
\usepackage{graphicx}

\begin{document}
\title{Mechanism of Tunneling\\ in Interacting Open Ultracold Few-Boson Systems}
\author{Axel U. J. Lode, Alexej I. Streltsov, Ofir E. Alon, and Lorenz S. Cederbaum}
\affiliation{Theoretische Chemie, Physikalisch-Chemisches Institut, Universit\"{a}t Heidelberg, Im Neuenheimer Feld 229, D-69120 Heidelberg, Germany}

\begin{abstract}
We investigate the mechanism in the tunneling dynamics of open ultracold few-boson systems by 
numerically solving the time-dependent few-boson Schr\"{o}dinger equation exactly. 
By starting from a weakly repulsive, initially coherent two-boson system we demonstrate 
that the decay dynamics incorporate fragmentation. 
The wavefunction of the tunneling state exhibits a pronounced dynamically-stable pattern 
which we explain by an analytical model. By studying more bosons and stronger 
interactions we arrive to the conclusion that the decay by tunneling is not a coherent process 
and exhibits a wealth of phenomena depending on the interaction between the particles.
\end{abstract}

\maketitle
One of the most basic and also paradigmatic phenomena in quantum mechanics is the
tunneling process in open systems on which ample studies have been reported 
(for a recent book, see \cite{complex}). Much attention has also been devoted to the tunneling 
of particles between potential wells (see, e.g., \cite{MKO,JJ_K}), 
but in the present work we concentrate on open systems, where the tunneling is from a 
well through a potential barrier into open space. Since Bose-Einstein 
condensates (BECs) \cite{RMD1,RMD2} have been realized, this unique state of matter has been 
used to model a wealth of physical phenomena in a controllable fashion \cite{arbpot}. 
BECs were made one-dimensional, see, e.g., \cite{1d1}, and there is progress towards 
the realization of ultracold few-boson systems \cite{chuu}. 
Such systems provide ideal real objects to investigate the mechanism behind the tunneling 
process in interacting systems and to arrive at a deeper understanding of the 
quantum dynamics of tunneling  in general. 

When studying tunneling the calculation is often reduced to that of a single particle. 
In the case of BECs the time-dependent Gross-Pitaevskii (TDGP) theory, 
which is known to be successful in many situations, 
naturally describes tunneling of the interacting bosons as the tunneling of a single effective boson. 
There are very few studies available which discuss tunneling in open systems beyond the Gross-Pitaevskii (GP) theory. 
For infinitely strong interacting bosons (Tonks-Girardeau gases), 
Del Campo \textit{et al.} \cite{campo} analytically treated the case of decay 
by tunneling through a Dirac-$\delta$ barrier. 
In \cite{schl-nim} the decay rates of a BEC in a double-barrier potential 
were computed by employing complex scaling to the GP and the best mean field methods. 
This investigation provided first indication that tunneling of an initially coherent 
system can lead to fragmentation, which is a known phenomenon of BECs confined 
to an external potential \cite{frag}. 
This finding has recently been substantiated in [12] where it 
is demonstrated that the TDGP theory fails to correctly
describe the tunneling of open few-boson systems even for weakly interacting particles. 
Interestingly, this failure is more severe for four than for two particles [12b]. 
Clearly,
the mechanism governing the physics of tunneling in open few-boson 
systems involves a dynamical loss of coherence of the wavefunction, 
a phenomenon which cannot be described by GP theory.

In this work we solve the time-dependent Schr\"{o}dinger equation (TDSE) numerically exactly, 
and unravel the mechanism of time-dependent tunneling in one dimension. 
We demonstrate fragmentation even for weak interactions. 
A particularly exciting stationary pattern of the tunneled fraction emerges which is 
analyzed by a simple analytical model. 
We start with two bosons and show that the dynamics in the case of weak interactions 
exhibit dynamical fragmentation: the density forms a dynamically-stable interference-like pattern, 
the momentum distribution exhibits two prominent peaks, 
and the largest occupation number decreases, i.e., the system becomes less condensed (coherent). 
All these features cannot be covered by TDGP theory. Increasing the interparticle repulsion, 
we verify that the density mimics fermionic behavior, 
expected in the limit of very strong interactions, see work by Girardeau \cite{TG}. 
Next, considering four particles we see that the dynamically-stable structure found for 
two bosons disbands and becomes time-dependent, 
but the phenomenon of dynamical fragmentation remains.

\begin{figure}
\includegraphics[width=0.5\textwidth]{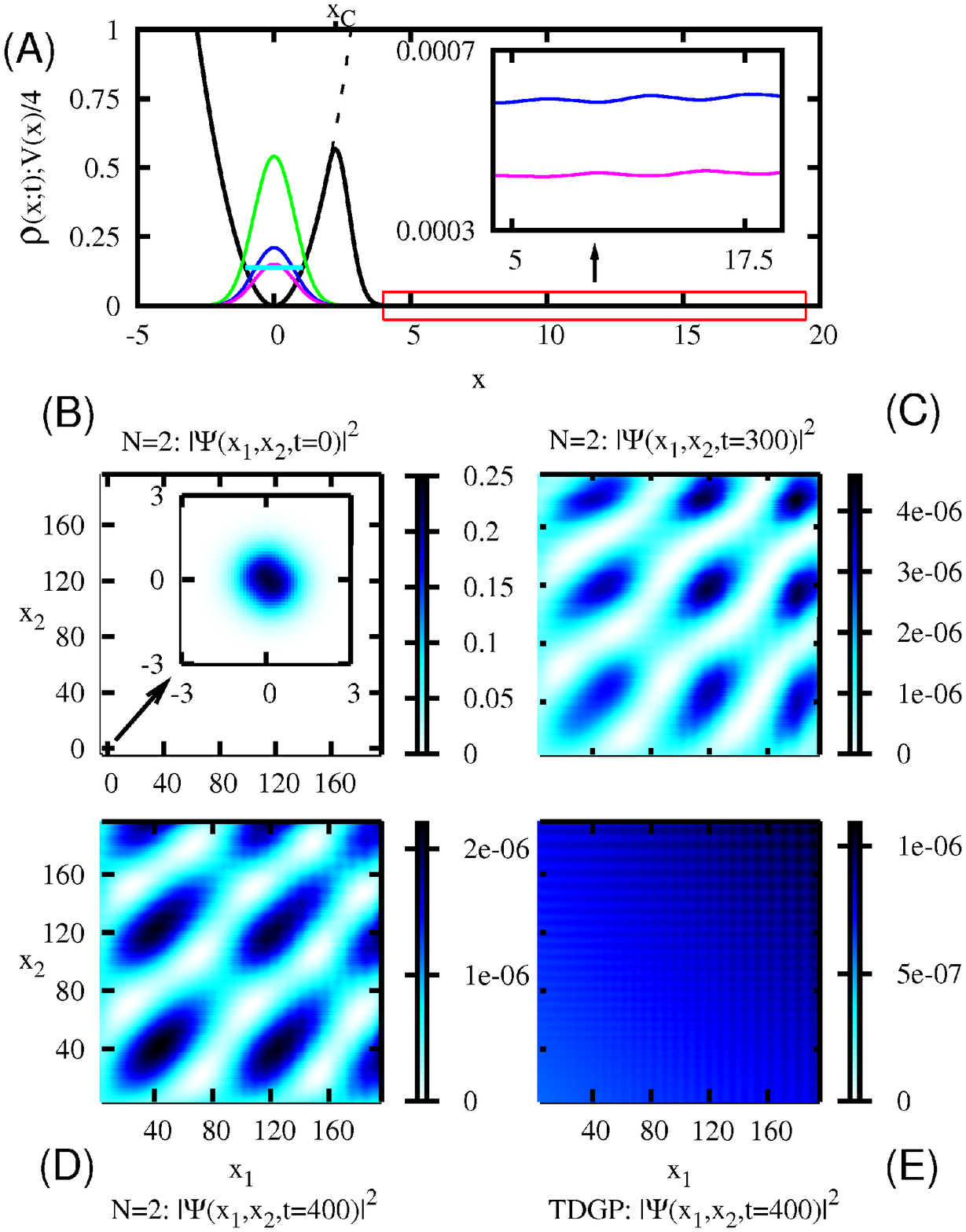}
\caption{(Color online) Tunneling dynamics of weakly interacting bosons: $\lambda=0.3$ and $N=2$ bosons.
(A): The initial parabolic trap is depicted by the black-dashed line. 
The potential used for the propagations is plotted as solid black line. 
The density at times $t=0,300,400$ is indicated by the green, 
blue and magenta line, respectively. 
The energy per particle is plotted as a turquoise horizontal line. 
The inset shows an enlargement of the density in the outside region. (B),(C),(D): 
The wavefunction (square) evolves from its initial Gaussian shape to a \textit{lobe-structure} 
which is persistent throughout the dynamics. (See text for details.) 
(E): For comparison, the TDGP shows minor oscillations 
and no \textit{lobe-structure.} All quantities shown are dimensionless.}
\label{2bden}
\end{figure}

We tackle the TDSE with the full many-body Hamiltonian, 
$\hat{H}=\sum_{j=1}^N \hat{h}(x_j) + \sum_{j<k}^N \hat{W}(x_j-x_k)$ in dimensionless units. 
Here the one-body Hamiltonian is given by $\hat{h}(x)= \hat{T}(x)+V(x)$. 
$\hat{T}$ is the kinetic energy operator $\hat{T}(x)=-\frac{1}{2}\partial^2_{x}$ and 
$V(x)$ is the potential energy operator, shown by the dashed and solid black lines in Fig.~\ref{2bden}A 
(for an analytical form see Ref.~[12]). To model the commonly used Dirac-$\delta$ two-body interaction 
we take $\hat{W}(x_j-x_k) = \lambda_0 \delta_{\sigma}(x_j-x_k)$ where $\delta_{\sigma}(x)$ is a normalized Gaussian, 
of width $\sigma=0.05$. $\lambda_0$ is the interaction strength. 
To emulate an open, infinite system we use a 
complex absorbing potential (CAP) \cite{CAP3,CAPnim}, 
which we add to the one-body Hamiltonian $\hat{h}(x)$. 
It absorbs the wavefunction from some point $x_S$ on and is 
adjusted such that it is essentially reflection free 
and perfectly absorbing (see \cite{CAP3}). 
We use the multi-configurational time-dependent Hartree (MCTDH) \cite{CPL,JCP} 
method in its parallel version \cite{michael} to solve the  
few-boson TDSE numerically exactly. 
This algorithm has already been successfully applied to the dynamics 
of few-boson systems (see, e.g., \cite{No1,sascha}). 
The parameters for convergence of the calculation with respect to the CAP, 
number of grid points 
and the number of orbitals are the same as in Ref.~[12]. 
In the present work, 
we enlarge the simulated region to $x\in[-6,226]$, 
and use $2048$ grid points. The CAP starts at $x_S=196$.

The most general quantity to describe 
a quantum mechanical system is the $N$-body density which is a 
function of as many variables as the system's constituents, $N$, plus one for the time. 
The simplest interacting system to consider is that
of two weakly interacting bosons.
Moreover, in this case we are able to \textit{visualize} the full
density, $\vert \Psi(x_1,x_2;t)\vert^2$, as a 3D-plot at a given time $t=t'$.
Thus, we prepare the ground-state of the weakly interacting two-boson system 
with $\lambda\equiv\lambda_0(N-1) = 0.3$ in a parabolic potential 
(see Fig.~\ref{2bden}A). This initial state is almost completely coherent, 
its largest occupation number $\rho_1(t=0)$ is $99.85\%$. 
Then we abruptly switch the parabolic trap to the 
open potential. 
Consequently, the wavepacket is decaying by tunneling 
through the barrier. 
In Figs.~\ref{2bden}~B-D we plot the density of the exact solution for the two-boson case 
with interparticle interaction strength $\lambda=0.3$ at times 
$t=0;300;400$, respectively. In Fig.~\ref{2bden}E the TDGP density at $t=400$ is depicted for comparison. 
First, we notice that a dynamically stable shape is formed and persists 
for long time on the scale of the half-life of the system. 
We refer to this shape of the density as the \textit{lobe-structure}. 
Second, the TDGP density, contrasting the exact solution, does not display such a lobe-structure. 
Several questions arise:  
Is the tunneling process a coherent process?
Does the initial almost fully coherent, weakly interacting system stay coherent 
throughout its time evolution? Surprisingly the answers are negative. 
The fact that the computed exact density differs drastically from the TDGP 
density allows us to interpret this difference as a loss of coherence. 
We recall that in the TDGP theory the wavefunction stays completely coherent at all times, i.e. $\rho_1(t)=100\%,\forall t$. 
Thus we have to ask: What is the mechanism behind the system's loss of its coherence and 
what makes the system form a dynamically stable shape in the density? 
Does this shape persist for more particles or stronger interactions? 
To answer the former question we will investigate the momentum distribution of the 
tunneling dynamics and to answer the latter we will subsequently look at the density 
of a strongly interacting two-boson system and at the two-body-density 
$\rho^{(2)}(x'_1=x_1,x'_2=x_2,x_1,x_2;t)\equiv \rho^{(2)}(x_1,x_2;t)$ \cite{RDM_K} 
of a four-boson system. 

\begin{figure}
\includegraphics[width=0.5\textwidth,angle=-90]{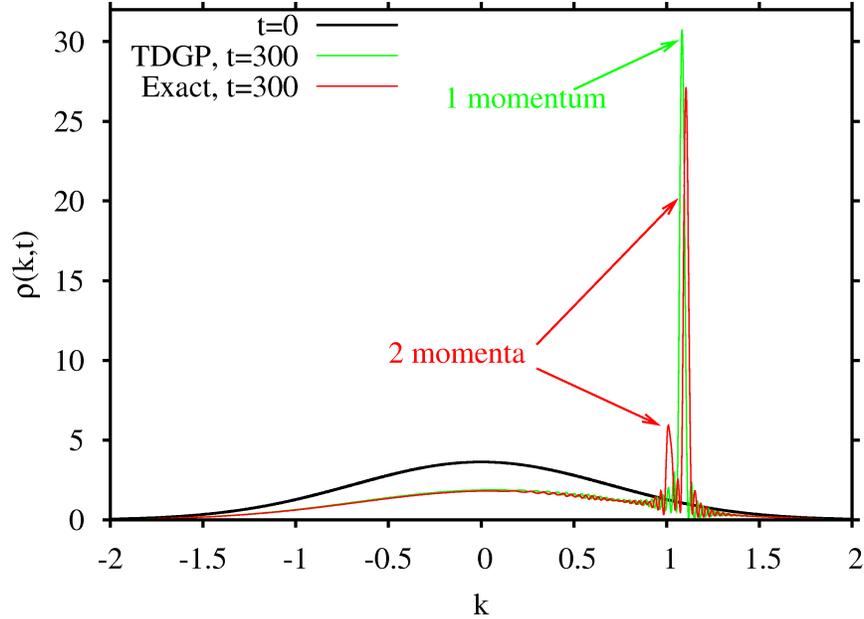}
\caption{(Color online) The momentum distribution for the times $t=0$ 
for the exact solution (black line) and $t=300$ for the exact solution 
(red line) and the TDGP solution (green line). 
The initial guesses are coherent, i.e., at $t=0$ the exact and the 
TDGP momentum distribution cannot be distinguished on the scale of the plot. 
The TDGP dynamics have a single-peaked momentum distribution. 
The exact momentum distribution reveals a two-peak structure. 
The intensity of the peak in the TDGP momentum distribution is roughly the sum 
of the intensities of the two exact peaks. All quantities shown are dimensionless.}
\label{momdist}
\end{figure}

To get a deeper insight on the many-body quantum dynamics we investigate the 
momentum distribution of the considered system. In Fig.~\ref{momdist} we compare the momentum distributions of the 
exact solution and the TDGP for the times $t=0,300$, respectively. 
We see that the momentum distributions of the initial guesses are similar and 
Gaussian-shaped, as it is expected for weakly interacting bosons in a parabolic trap. 
For the time-evolution, in turn, 
we get a persistingly prominent \textit{two-peak} structure on a Gaussian background for the exact solution and 
a persistingly prominent \textit{single-peak} structure for the TDGP dynamics. 
We attribute the peak-structure to the part of the density running away from the well. 
Thus, we can conclude that the single momentum dominated picture of the dynamics is caused by the 
constraint that the system stays in a \textit{single} quantum mechanical 
state, i.e., is fully coherent. 
But the exact density shows two peaks in the momentum distribution. Hence, we can suppose that the system 
occupies two states, exhibiting two distinct momenta. 
A bosonic system occupying two quantum mechanical states is fragmented - 
this allows us to state that the loss of the initial coherence is caused by the tendency 
of the system to fragment in the course of the tunneling dynamics outside the barrier. 

\begin{figure}
\includegraphics[width=0.5\textwidth]{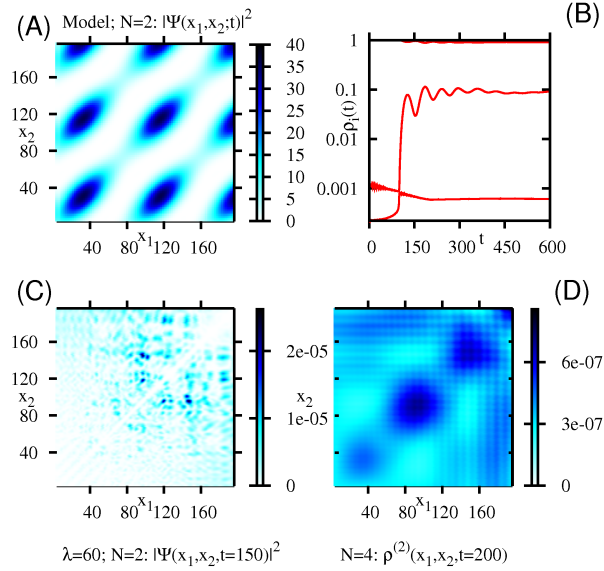}
\caption{(Color online) Self-interference and fragmentation 
is the mechanism governing the tunneling dynamics. 
(A): The model introduced in Eq.~(\ref{model}) for the wavefunction (square) 
is plotted for a \textit{fully} 
fragmented outer wavefunction. Note the similarity to the dynamically-stable 
structure depicted in Figs.~\ref{2bden}(C) and (D). 
(B): The time-evolution of the occupation numbers $\lbrace \rho_i(t); i=1,2,3 \rbrace$ 
(normalized to $1\;\forall t$) for $N=2$ weakly interacting bosons indicate the dynamical fragmentation of the system. 
(C): The wavefunction (square) $\vert \Psi(x_1,x_2; t=150) \vert^2$ 
for a rather strong interaction of the $2$ bosons. The diagonal, $x_1=x_2$, tends to $0$. 
We show only one specific time, $t=150$, here, but in general the off-diagonal part 
shows short-range oscillations and strong time-dependency. 
(D): The two-body density, $\rho^{(2)}(x_1,x_2;t=200)$, of $N=4$ weakly interacting bosons. 
Although we depict only $t=200$, let us state that $\rho^{(2)}$ is time-dependent 
and forms no dynamically-stable pattern. See text for discussion. All quantities shown are dimensionless.}
\label{perspective}
\end{figure}

We observe the lobe-structure in the density for a whole range of interactions, $0<\lambda_0 \lesssim 2$. 
We have also found that the distance of the peaks increases 
with the increase of repulsion, whereas the length scale of the lobe-structure decreases. 
These observations motivated us to formulate an analytical model 
for the wavefunction. We assume that in the \textit{outside region}, i.e., at $x>x_C$, 
the wavefunction is in general constituted by plane waves. 
The repulsion forces the bosons to occupy two plane waves, 
i.e., $\phi_{j}(x,t)=(1-\Theta_{x_C}) a_{in,j}(t) \phi_{in,j}(x) +\Theta_{x_C} a_{j}(t) e^{ik_{j}x};j=1,2$.
The $\phi_{in,j}$ are the ground and first excited state of a single particle in a harmonic potential. 
The coefficients $a_{in,j}(t),a_j(t)$ describe the time-dependent occupations of 
the inner parts, $\phi_{in,j}(x,t)$, and the plane waves, respectively. 
The Heavyside function, 
$\Theta_{x_C}\equiv\Theta(x - x_C)$, starts at the barrier, $x_C=2.25$, and the $k_j$ are the 
positions of the two peaks occurring in the momentum distribution of the 
tunneling dynamics. 
Assuming further that the coefficients $a_j(t)$ are real we can construct the full density from these orbitals:
\begin{eqnarray}
 \rho^{(2)}(x_1,x_2;t)&=& (1-\Theta_{x_C})\rho_{in}^{(2)}(x_1,x_2;t) \nonumber\\
 &+& \Theta_{x_C} \Bigg\lbrace A(t) \cos \left[\left(k_2-k_1\right)x_1\right] \cos\left[\left(k_2-k_1\right)x_2\right] \label{model}\\
 &+& B(t)\cos\left[ \left(k_2-k_1\right) (x_1+x_2)\right]  +C(t)  \Bigg\rbrace .\nonumber 
\end{eqnarray}
Here the coefficients $A(t)$, $B(t)$ and $C(t)$ collect the dependency on the coefficients 
$a_j(t)$, and $\rho^{(2)}_{in}$ is a symmetrized product of the $\phi_{in,j}$. 
In Fig.~\ref{perspective} we plot $\rho^{(2)}(x_1,x_2;t)$ and take $(k_2-k_1)=0.15$ from Fig.~\ref{momdist} 
for the distance of the momentum peaks and $A(t)\equiv B(t)$, $C(t)=0$. 
We note here that this choice of $A,B$ and $C$ corresponds to a \textit{fully} fragmented wavefunction 
outside the well. Comparing Fig.~\ref{perspective}A with Fig.~\ref{2bden}B and C, 
we find that our model reproduces very well all key features found.
The lobe-structure is stable in time for a range of interactions. 
Another nice feature of the model is 
that it also reproduces the TDGP results as a limiting case. 
Indeed, if we consider a single $\phi_j(x,t)$ [take $k_1=k_2$ in Eq.~(\ref{model})], 
we obtain the single value of $A(t) + B(t) + C(t)$ in the outside region. 
So, the non-oscillatory, spatially-independent behavior of the 
TDGP density in Fig.~\ref{2bden}E is reproduced. 
We conclude that the mechanism behind 
the loss of coherence of the system is that 
the tunneled density dynamically fragments and occupies mainly the permanents formed by two plane waves given in Eq.~(\ref{model}). 
Thus, the lobe-structure is a \textit{self-interference} pattern. 
Clearly this process cannot be described at all by the TDGP theory. 

The occurring fragmentation can also be observed by 
looking at the time evolution of the $i=1,...,M$ renormalized natural occupation numbers $\rho_i(t)$.
In Fig.~\ref{perspective}B we plot their time evolution. 
The fragmentation appears as an increase of $\rho_2(t)$ to about $10\%$ 
accompanied by a decrease of $\rho_1(t)$ to around $90\%$. 
The other occupation numbers, $\rho_{i>2}(t)$, stay very small at all times.
This describes the loss of coherence in the tunneling dynamics from another viewpoint.
Let us analyze the dynamics of the occupation numbers a little closer: 
The initial time evolution is accompanied by oscillations of 
$\lbrace \rho_i(t); i=1,2 \rbrace$. 
From our model consideration we can conclude that the relative occupancy, 
$a_1(t)/a_2(t)$, of the plane waves constituting the outside 
fraction of the wavefunction does not change much in time - 
that is why we find a dynamically-stable density. 
Therefore, the mentioned oscillations are caused by the density localized {\it in} the well.
We term the oscillations \textit{beating}. 
We find that the frequency of the beating, $\omega$, is related 
to momentum by $\omega=(k_1-k_2)^2/2$, where $\vert k_1-k_2\vert$ is the spacing 
of the two peaks in the momentum distribution in Fig.~\ref{momdist}. 
Thus, the beating and the appearance of the two 
peaks in the momentum distribution are intimately related.

We analyze the dynamics of two more strongly repelling bosons, 
to see whether and how the above discussed mechanism of tunneling dynamics is altered.
When we increase the inter-boson repulsion further ($\lambda \gtrsim 2$), 
the dynamically-stable structure dissolves and becomes time-dependent. 
According to the reasoning of our model this happens when more than 
two states outside the well become populated. 
Obviously, the process of loss of coherence intensifies in this case as the wavefunction 
occupying more than two states is less coherent than the one occupying two. 
When we increase the interaction strength to the limit of very strong repulsion, 
e.g., $\lambda \gtrsim 60$, the density starts to mimic the one 
in the fermionization limit (see \cite{TG}). 
In Fig.~\ref{perspective}C we plot the computed density for $\lambda=60$ and $N=2$ at $t=150$. 
It is clearly seen that the density's diagonal becomes $0$, 
i.e., $\Psi(x_1=x_2;t)=0$. 
The off-diagonal shows strong dependence on time.

Let us see now how the found mechanism behind the tunneling process 
develops when increasing the particle number. 
We investigate a four-boson system with an equal mean field parameter $\lambda=\lambda_0(N-1)=0.3$. 
We recall that 
all systems with the same $\lambda$ reveal identical TDGP dynamics, irrespective to changes in the number of bosons, $N$. 
Now, i.e., for $N=4$, we consider the diagonal part of the reduced two-body density matrix, 
$\rho^{(2)}(x_1,x_2;t)$, in order to be able to compare with $N=2$, 
where the density $\vert \Psi(x_1,x_2;t) \vert^2$ 
is equal to $\rho^{(2)}(x_1,x_2;t)$. 
In Fig.~\ref{perspective}D we plot $\rho^{(2)}$ for $N=4$ at $t=200$.
The shape of the density exhibits time-dependency in the outside region. This 
indicates that adding more bosons to the system also forces it 
to occupy more than two states during the dynamics. 
We find that the dynamically-stable self-interference pattern 
is not persisting as the process of loss of coherence intensifies when we increase $N$. 
This explains why the survival probability computed on the many-body 
level deviates from the TDGP results
even more for four than for two bosons [12b]. 

Let us conclude. We have identified the mechanism governing the 
tunneling dynamics in open ultracold bosonic systems based on an exact numerical solution of the 
TDSE. The results were compared to the TDGP approximation, which is widely and successfully used, 
but is not applicable to the scenario studied in this work. 
We explain the considerable deviations of the exact solution from the TDGP 
dynamics by the fragmentation of the wavefunction, 
occupying more states in the outside region. 
For the case of two weakly interacting bosons we observe a lobe-structure which is dynamically-stable. 
A simple analytical model of a two-fold fragmented state formed by two plane waves with different momenta explains this lobe-structure. 

For stronger interactions 
for two bosons and for a larger number of bosons the 
dynamically-stable structure breaks and becomes time-dependent, indicating thereby that the loss of coherence intensifies. 
The tunneling of a few interacting bosons is not 
a coherent process and the interference 
effects taking place in the outgoing fraction of the wavepacket are in this case a signature of fragmentation. 

\begin{acknowledgments}
Fruitful discussions with K. Sakmann, M. Brill, and
H.-D. Meyer, financial support by the DFG, and computing time at the bwGRiD are gratefully acknowledged.
\end{acknowledgments}

\clearpage

\clearpage

\clearpage
\clearpage

\end{document}